\documentclass[aps,prl,showpacs,twocolumn,floatfix,nofootinbib,superscriptaddress]{revtex4-1} 

\usepackage{amsmath,amssymb,amsfonts,graphicx}
\usepackage{bm}
\usepackage{slashed}
\usepackage{subfigure}
\usepackage[svgnames]{xcolor}
\usepackage[section]{placeins}
\usepackage{xspace}

\def\kt{k_T}

\def\LEPT{{\tt LEPTO} }

\def\vfR{$\varphi_R$\xspace}
\def\fR{\varphi_R}
\def\fS{\varphi_S}
\def\vfT{$\varphi_T$\xspace}
\def\fT{\varphi_T}
\def\kp{\vect{k}_T}

\def\Pp{{\vect{P}_T}}
\def\Pp#1{\vect{P}_{#1T}}
\newcommand{\vect}[1]{\boldsymbol{#1}}
\newcommand{\bfk}{\vect{k}}
\newcommand{\bfq}{\vect{q}}
\newcommand{\bfP}{\vect{P}}
\newcommand{\bfS}{\vect{S}_{_T}}

\newcommand{\bfp}{\vect{p}}

\newcommand{\al}[1]{\begin{align} #1 \end{align}}
\newcommand{\non}{\nonumber}
\newcommand{\vf}{\varphi}

\newcommand{\Ge}{\mathrm{GeV}}
\newcommand{\Gs}{\mathrm{GeV}^2}
\newcommand{\ImS}{0.66\columnwidth}
\begin{document}

\title{Sivers Effect in Two-Hadron Electroproduction}

\preprint{ADP-14-9/T867}

\author{Aram~Kotzinian}
\affiliation{Yerevan Physics Institute,
2 Alikhanyan Brothers St.,
375036 Yerevan, Armenia
}
\affiliation{INFN, Sezione di Torino, 10125 Torino, Italy
}

\author{Hrayr~H.~Matevosyan}
\email{hrayr.matevosyan@adelaide.edu.au}
\affiliation{ARC Centre of Excellence for Particle Physics at the Tera-scale,\\ 
and CSSM, School of Chemistry and Physics, \\
The University of Adelaide, Adelaide SA 5005, Australia
\\ http://www.physics.adelaide.edu.au/cssm
}

\author{Anthony~W.~Thomas}
\affiliation{ARC Centre of Excellence for Particle Physics at the Tera-scale,\\     
and CSSM, School of Chemistry and Physics, \\
The University of Adelaide, Adelaide SA 5005, Australia
\\ http://www.physics.adelaide.edu.au/cssm
}

\begin{abstract}
The Sivers effect in single hadron semi-inclusive deep inelastic scattering (DIS) on a transversely polarized nucleon describes the modulation of the cross section with the sine of the azimuthal angle between the produced hadron's transverse momentum and the nucleon spin ($\vect{P}_h$ and $\fS$ respectively). This effect is attributed to the so-called Sivers parton distribution function of the nucleon. We employ a simple phenomenological parton model to derive the relevant cross section for two-hadron production in semi-inclusive deep inelastic scattering including the Sivers effect. We show that the Sivers effect can be observed in such process as sine modulations  involving the azimuthal angles \vfT~and \vfR~of both the total and the relative transverse momenta of the hadron pair. The existence of the modulation with respect to~\vfR~is new. Finally, we employ a modified version of the \LEPT Monte Carlo event generator that includes the Sivers effect to estimate the size of single spin asymmetries corresponding to these modulations. We show that~$\sin(\fR-\fS)$ modulations can be significant, especially if we impose asymmetric cuts on the momenta of the hadrons in the pairs.
\end{abstract}


\date{\today}                                           

\maketitle

 The exploration of the structure of the nucleon is one of the most important issues in medium energy nuclear physics. Experiments such as those at JLab, HERMES, COMPASS and the upcoming JLab12 and EIC programs are exploring the three dimensional distribution of the parton's momentum and spin inside the nucleon. One of the exciting phenomena to be investigated here is the so-called Sivers effect~\cite{Sivers:1989cc}, describing the correlation of the transverse momentum of the unpolarized partons with the transverse spin of the nucleon, which is quantified by the Sivers parton distribution function (PDF). The Sivers PDF can be accessed in both Drell-Yan leptoproduction and the lepton-nucleon semi-inclusive deep inelastic scattering (SIDIS) process with a transversely polarized target and a single detected hadron~\cite{Collins:2002kn}. In SIDIS, the Sivers function can be accessed by measuring the single spin asymmetry (SSA) modulated with respect to the sine of the difference of the azimuthal angle of the produced hadron and the nucleon's transverse spin vector~\cite{Anselmino:2005nn}. Here the SSA is a ratio of a convolution of the Sivers PDF with the unpolarized fragmentation function (FF) to that of the unpolarized PDF and FF. These asymmetries have been measured in HERMES~\cite{Airapetian:2009ae}, COMPASS~\cite{Adolph:2012sp}, and JLab HALL A~\cite{Qian:2011py} experiments.

 In this Letter we propose a new approach for measuring the Sivers function in the SIDIS process involving a transversely polarized target and two detected hadrons. We calculate the cross section of such a process by assuming simple parton-model inspired functional forms for both the unpolarized and Sivers PDFs, as well as the unpolarized dihadron FF (DiFF). Our results show that the relevant terms of the cross section involving the Sivers PDF do not vanish. There are two corresponding SSAs, one with respect to $\sin(\fR-\fS)$, and another with respect to $\sin(\fT-\fS)$. Here \vfT, \vfR and $\fS$ are the  azimuthal angles of the total and the relative transverse momenta of the hadron pair, and the transverse spin of the nucleon. Finally, we use the \LEPT Monte Carlo event generator~\cite{Ingelman:1996mq}, modified to include the Sivers effect~\cite{Kotzinian:2005zs,Kotzinian:2005zg}, to show that such modulations are, indeed, generated in SIDIS in the kinematical region of the COMPASS experiment.  In contrast, the leading twist expression for the cross section of the two-hadron SIDIS presented in Ref.~\cite{Bianconi:1999cd} contains only the $\sin(\fT-\fS)$ modulation term. The $\sin(\fR-\fS)$ modulation term is absent there, as well as in the subleading twist expression of Ref.~\cite{Bacchetta:2003vn} for the same cross section, when integrated over the total transverse momentum of the pair (note that our definition of $\fR$ is different  than those in Refs.~\cite{Bianconi:1999cd,Bacchetta:2003vn}).
\begin{figure}[tbh]
\begin{center}
\includegraphics[width=0.9\columnwidth]{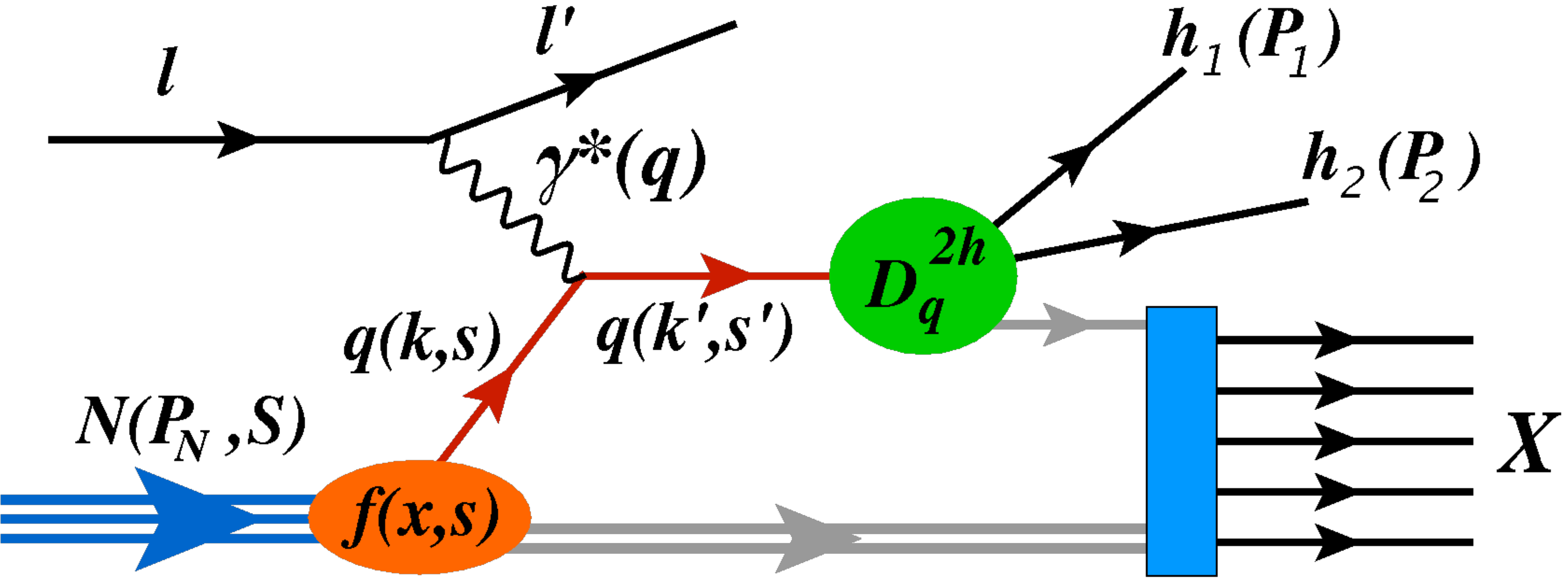}
\caption{The leading order diagram for two-hadron production in the current fragmentation region of SIDIS.}
\label{FIG_DIHADRON_CFR}
\end{center}
\end{figure}

%
{\it The Sivers effect in two-hadron production.}---Let us consider the process of two-hadron electroproduction in SIDIS, depicted schematically in Fig.~\ref{FIG_DIHADRON_CFR}
\al{
\label{EQ_2H_SIDIS}
\ell (l ) + N({P_N},S) \to \ell (l') + h_1(P_1) + h_2(P_2) + X\,.
}
Here, the lepton with momentum $l$ scatters off a nucleon with momentum $P_N$ and spin $S$, and the scattered lepton with momentum $l'$ and two hadrons with momenta $P_1$ and $P_2$ are detected in the final state. We adopt the $\gamma^*-N$ center of mass frame, where the virtual photon and the proton collide along the $z$ axis with momenta $\bfq$ and $-\vect{P}_N$ respectively, and the leptonic plane (defined by $l$ and $l'$) coincides with the $x$-$z$ plane. The conventional SIDIS variables are defined as
\al{
&
q=l-l',\,  Q^2 = -q^2,\, x = \frac {Q^2}{2P_N \cdot q},
\\&\non
\, y = \frac{P_N \cdot q}{P_N \cdot l},\, z_i = \frac{P_N \cdot P_i}{P_N \cdot q}.
\label{EQ_KIN}
}

The Sivers effect arises because of the correlation between the active quark transverse momentum, $\vect{k}_T$, and the transverse polarization of the nucleon, $\mathbf{S}_T$. The scalar quantity describing this correlation can be written as $\hat{\vect{P}}_N \cdot [\bfk_T \times \bfS]=[\bfS \times \bfk_T]_3=S_T k_T \sin\left(\vf_k-\fS\right)$, with $\hat{\vect{P}}_N$ standing for unit vector in the direction of $\vect{P}_N$ and subscript 3 denotes the $z$ component of the vector in the $\gamma^*-N$ c.m. frame. The quark transverse-momentum-dependent distribution in the polarized nucleon can be expressed as
\al{
f_\uparrow^q(x,\vect{k}_T)=f_1^q(x,k_T)+\frac{[\bfS \times \bfk_T]_3}{M}f_{1T}^{ \bot q}(x,k_T).
}

The factorization theorem tells us that the cross section for the process in Eq.~(\ref{EQ_2H_SIDIS}) involves the same PDFs as for single hadron production, but instead of the quark single hadron FF, now a two-hadron fragmentation function (DiFF), $D_{q}^{h_1,h_2}$, is involved (see Fig.~\ref{FIG_DIHADRON_CFR}). Note, that, in this case, the FFs depend on the light-cone momentum fractions of the hadrons, $z_1$ and $z_2$, as well as their transverse momenta with respect to the fragmenting quark's momentum, $\vect{p}_{1\perp}$ and $\vect{p}_{2\perp}$ (and $Q^2$). Keeping only the terms relevant for Sivers effect, the cross section for the process Eq.~(\ref{EQ_2H_SIDIS}) at the leading twist can be expressed as
\al{
\non
& \frac{d\sigma}{dx\, d{Q^2}\, d{\fS}\, dz_1\, dz_2\, d^2\Pp{1}\, d^2\Pp{2}} = C(x,Q^2) \left({\sigma_U} + {\sigma_{S}}\right),
\\ \non
& \sigma_U = \sum_q e_q^2 \int d^2 \kp \ f_1^q\ D_q^{h_1h_2},
\\
\label{EQ_2H_SIG_SIV}
&\sigma_{S} = \sum_q e_q^2 {\int {{d^2}{\kp}} 
\frac{[\bfS \times \bfk_T]_3}{M} f_{1T}^{ \bot q}\ D_{1q}^{h_1,h_2}},
}
where $C(x,Q^2)$ is a kinematic factor. $\bfP_{1T}$ and $\bfP_{2T}$ are the transverse momenta of the produced hadrons with respect to the $z$ axis. Here, we employ the leading order approximations $\bfp_{1\perp} \approx \bfP_{1T}-z_1\bfk_T$ and $\bfp_{2\perp} \approx \bfP_{2T}-z_2\bfk_T$. Using rotational and parity invariance, it is easy to show that for the most general transverse momentum dependence of the Sivers function and DiFF, the spin dependent part of the cross section contains two "Sivers-like" azimuthal modulations
\al{
\non
\sigma_{S} = S_T\left[\sigma_1\frac{P_{1T}}{M}\sin(\vf_1-\fS)+
\sigma_2\frac{P_{2T}}{M}\sin(\vf_2-\fS)\right],
}
where $\vf_1$ and $\vf_2$ are the azimuthal angles of the first and second hadron and $\sigma_{S}, \sigma_1$, and $\sigma_2$ depend on $x,Q^2,z_1,z_2,P_{1T},P_{2T},\vect{P}_{1T}\cdot\vect{P}_{1T}$. In general, all three structure functions $\sigma_U, \sigma_1$, and $\sigma_2$ depend on $\cos(\vf_1-\vf_2)$,  and a rather strong (back-to-back) correlation between the transverse momenta of the hadrons was observed in the process under consideration for an unpolarized target~\cite{Arneodo:1986yc}. Thus we have three types of azimuthal correlations: $\vf_1-\vf_2$, $\vf_1-\fS$, and $\vf_2-\fS$. The dependence of the cross section arising from the Sivers effect on the last two correlations is explicitly known from general principles.

In this Letter we adopt the common, simplifying assumption for both the parton densities and the fragmentation functions of the usual factorization between the intrinsic transverse momentum and the energy fraction dependence, with a Gaussian transverse momentum dependence, that is
\al{
f_{1q}(x,k_T) = f_{1q}(x) \, \frac{1}{\pi \mu_0^2} \, e^{-{\kt^2}/{\mu_0^2}}.
\label{EQ_UNPOL_PDF}
}
Here, $\mu_0^2=\langle \kt^2 \rangle$ is the mean value of the intrinsic transverse momentum of quark in the nucleon. For the Sivers function, we adopt the parametrization of Ref.~\cite{Anselmino:2005nn}
\al{
f_{1T}^{ \bot q}(x,k_T) = f_{1T}^{ \bot q}(x) \, \frac{1}{\pi \mu_{0,S}^2}
e^{-{\kt^2}/\mu_{0,S}^2} \,.
\label{EQ_SIV_PDF}
}

Further, we adopt a simple parametric form for the unpolarized DiFF
\al{
\label{EQ_DiFF_PARAM}
&D_{1q}^{h_1h_2}\left(z_1,z_2,p_{1\perp},p_{2\perp},\vect{p}_{1\perp}\cdot\vect{p}_{2\perp}\right)=
\\&\non
D_{1q}^{h_1h_2}(z_1,z_2)\frac{1}{\pi^2 \mu_1^2 \mu_2^2} \, e^{-p_{1\perp}^2/\mu_1^2-p_{2\perp}^2/\mu_2^2}
\left(1+c\,\vect{p}_{1\perp} \cdot \vect{p}_{2\perp}\right),
}
where the term $c \,\vect{p}_{1\perp} \cdot \vect{p}_{2\perp}$ takes into account the transverse momenta correlation in fragmentation. The parameters $\mu_1^2$, $\mu_2^2$, and $c$ in general can depend on the flavor of the fragmenting quark $q$, the type of the produced hadrons, and the kinematic variables $Q^2, z_{1}$, and $z_2$~\cite{Matevosyan:2011vj}.  

With this choice of DiFF, the integration over intrinsic transverse momentum in Eq.~(\ref{EQ_2H_SIG_SIV}) can be performed explicitly
\al{
\label{Eq:2h-str-fun}
&\sigma_U = \sum\limits_q e_q^2 f_1^q(x)D_{1q}^{h_1h_2}(z_1,z_2)C_0^{h_1h_2},
\\&\non
\sigma_1 = \sum\limits_q e_q^2 f_{1T}^{\perp q}(x)D_{1q}^{h_1h_2}(z_1,z_2)C_1^{h_1h_2},
\\&\non
\sigma_2= \sum\limits_q e_q^2 f_{1T}^{\perp q}(x)D_{1q}^{h_1h_2}(z_1,z_2)C_2^{h_1h_2},
}
where the explicit expressions for the $C_{i}^{h_1h_2}$ terms are presented in our forthcoming article~\cite{Kotzinian:2014siv}.

In the recent studies of two-hadron production, another choice of independent transverse momentum variables are often used~\cite{Bianconi:1999cd}: the total transverse momentum of the hadron pair, $\vect{P}_T=\vect{P}_{1T}+\vect{P}_{2T}$, and the half  of the relative transverse momentum of the two hadrons, $\vect{R}=\left(\vect{P}_{1T}-\vect{P}_{2T}\right)/2$, with corresponding azimuthal angles $\vf_T$ and $\vf_R$. It is easy to see that the modulations in terms of azimuthal angles of each hadron are related to the modulation in terms of  $\vf_T$ and $\vf_R$
\al{
\non
\frac{d\sigma}{d^2\vect{T}\, d^2\vect{R}\,} = C(x,Q^2)  &\left(\sigma_U + 
S_T \left[\sigma_T\frac{T}{M}\sin(\vf_T-\fS) \right.\right.
\\&
\left.\left.+ \sigma_R\frac{R}{M}\sin(\vf_R-\fS)\right]\right),
\label{EQ_2H_X_SEC_RT}
}
where $\sigma_T = \frac{1}{2}\left(\sigma_1+\sigma_2\right)$,  $\sigma_R = \sigma_1-\sigma_2$, and we omitted the dependence of the cross section on the rest of the variables for brevity. Here, the structure functions $\sigma_U, \sigma_T$ and $\sigma_R$ depend on $x,Q^2,z_1,z_2,P_T,R$n and $\vect{P}_T\cdot\vect{R}=P_T R\cos(\vf_T-\vf_R)$. The explicit expressions for these functions are presented in Ref.~\cite{Kotzinian:2014siv}. Nevertheless, we note that, in general, $\sigma_R \neq 0$. This can be ensured, for example, by choosing asymmetric cuts on the minimum values of $z_1$ and $z_2$. 

It is interesting to consider Eq.~(\ref{EQ_2H_X_SEC_RT}) after integrating over  the azimuthal angle of the relative or total transverse momentum, respectively,
\al{
&
\frac{d\sigma}{d^2\vect{T} RdR} \propto \sigma_{U}^0 +
S_T \left(\frac{T}{M}\sigma_{T}^0+\frac{R}{2M}\sigma_{R}^1 \right)\sin(\vf_T-\fS),
\label{EQ_2H_X_SEC_INT_R}
\\ &
\frac{d\sigma }{TdT d^2\vect{R}} \propto \sigma_{U}^0 +
S_T \left(\frac{T}{2M}\sigma_{T}^1+\frac{R}{M}\sigma_{R}^0 \right)\sin(\vf_R-\fS),
\label{EQ_2H_X_SEC_INT_T}
}
where $\sigma_{U}^i, \sigma_{T}^i$, and $\sigma_{R}^i$ are the zeroth ($i=0$) and the first ($i=1$) harmonics of the $\cos(\vf_1-\vf_2)$ expansions of the corresponding structure functions. Again, it is shown in Ref.~\cite{Kotzinian:2014siv} that, in general, the Sivers effect is nonzero in Eqs.~(\ref{EQ_2H_X_SEC_INT_R},\ref{EQ_2H_X_SEC_INT_T}).

\begin{figure}[ht]
\centering 
\subfigure[] {
\includegraphics[width=\ImS]{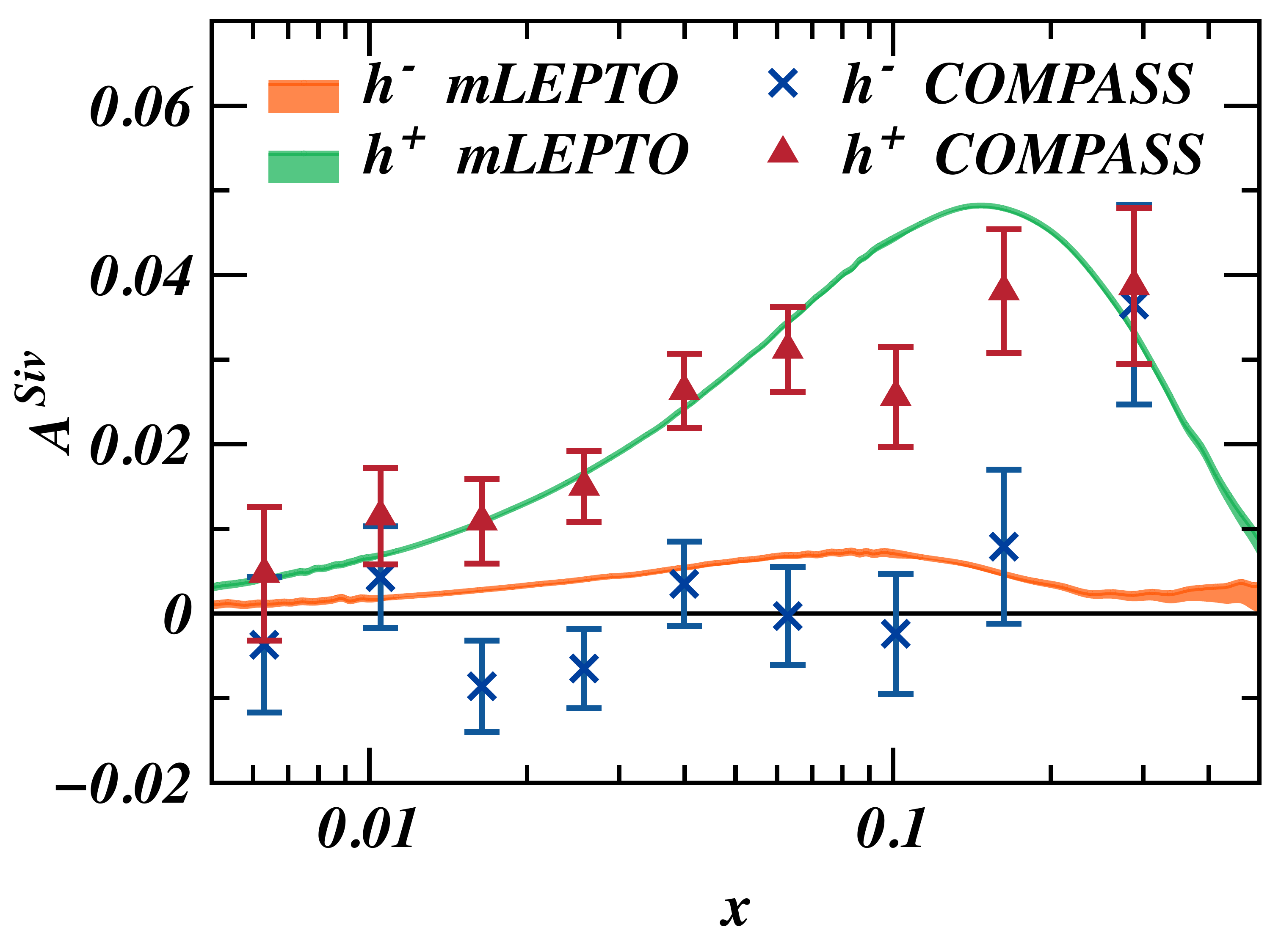}}
\\\vspace{-0.2cm}
\subfigure[] {
\includegraphics[width=\ImS]{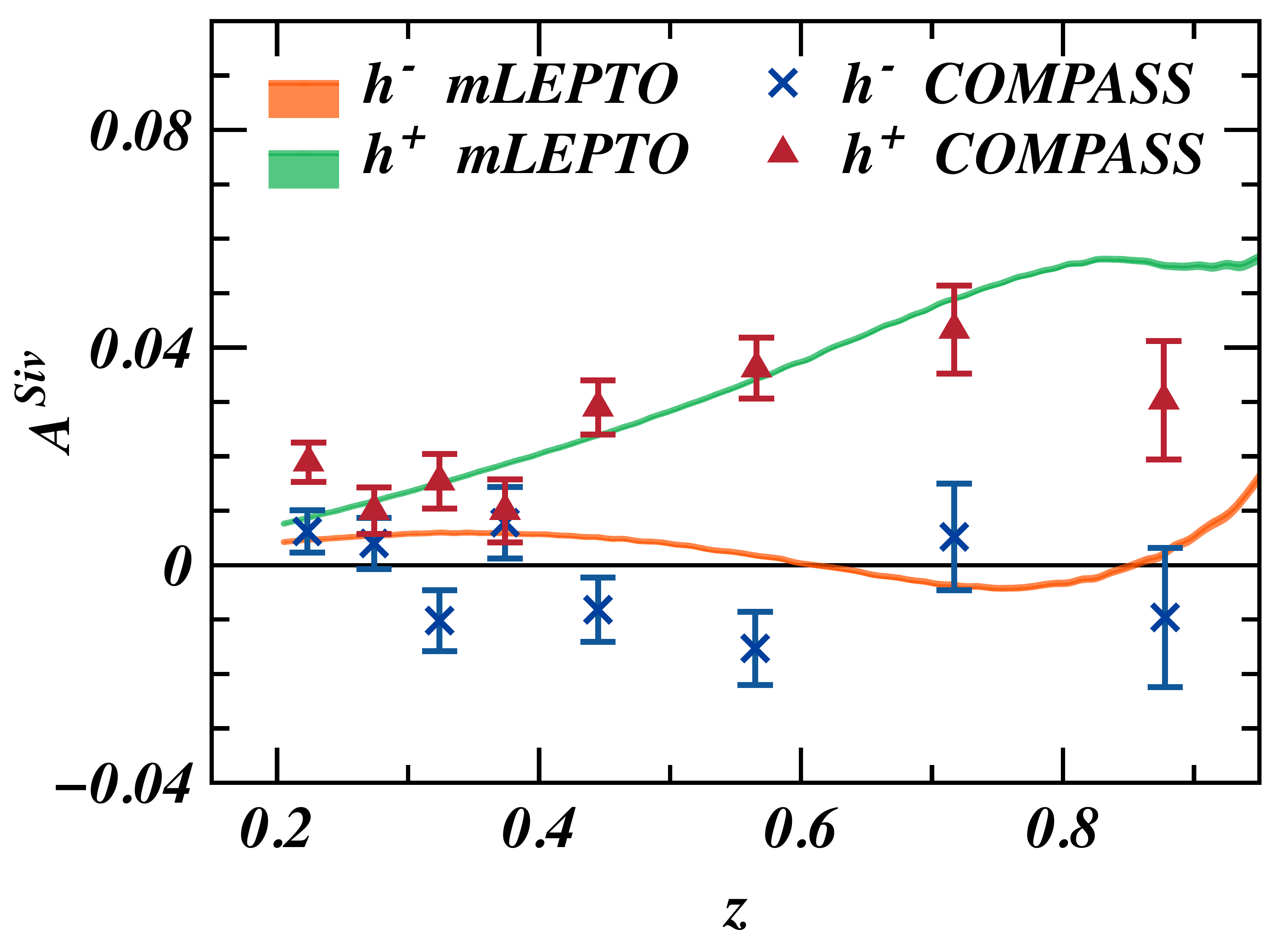}
}
\\\vspace{-0.2cm}
\subfigure[] {
\includegraphics[width=\ImS]{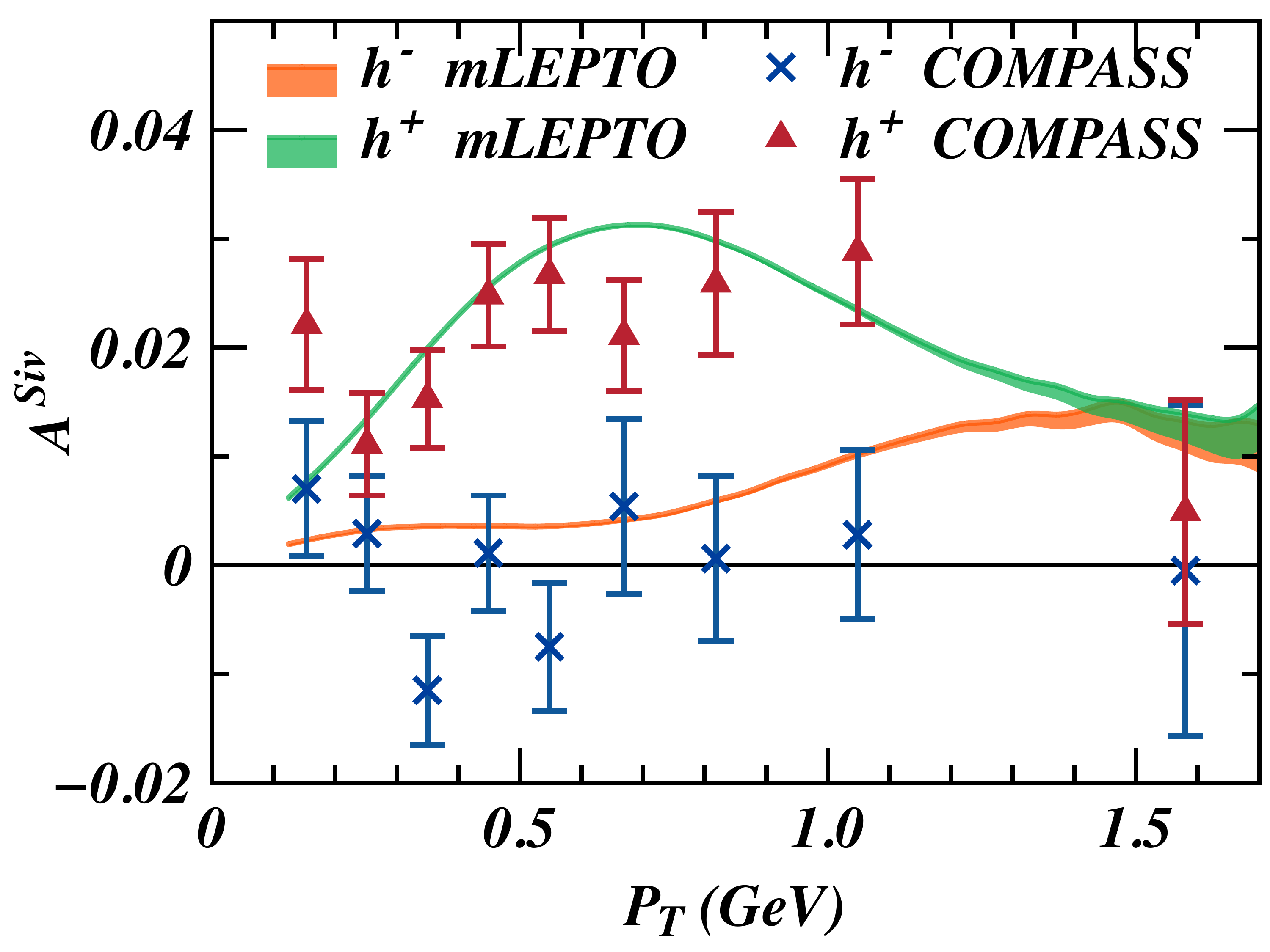}
}
\\\vspace{-0.2cm}
\caption{COMPASS results for Sivers asymmetry in a charged (triangles for positive and crosses for negative) hadron production off proton target, compared to those from mLEPTO (lines), for  $x$ (a), $z$ (b), and  $P_T$ (c) dependencies. The width of each line is larger than the statistical accuracy of our simulations and doesn't include the uncertainties of the PDFs.}
\label{PLOT_SIV_1H}
\end{figure}

{\it Modified \LEPT (mLEPTO) including the Sivers effect at leading order.}---The \LEPT unpolarized event generator~\cite{Ingelman:1996mq} is an invaluable tool  for studying both inclusive and semi-inclusive DIS reactions. However, in the leading order of QCD, this (and any other) generator does not simulate the experimentally observed azimuthal asymmetries for hadron production on both unpolarized  and polarized targets. We have modified the \LEPT code (mLEPTO) to include both Cahn and Sivers azimuthal modulations of the transverse momentum of the active quark before hard scattering and hadronization~\cite{Kotzinian:2005zs,Kotzinian:2005zg}. We employ mLEPTO in this work to first describe the SSAs for the Sivers effect in the single hadron SIDIS process as a validation of the event generator. Then, we use it to study the Sivers effect induced SSAs in two-hadron SIDIS, which is the focus of this Letter. We note that the hadronization in the Lund model, implemented in \LEPT, is  different from the independent FFs of parton model factorization theorem. Nevertheless, these two approaches produce similar results for the current fragmentation region of COMPASS kinematics~\cite{Kotzinian:2004xq}.
 
 We generated $10^{11}$ DIS events in mLPETO using the kinematics of the COMPASS experiment~\cite{Adolph:2012sp}: the energy of scattering muons $E_\mu=160~\Ge$,  $Q^2>1~\Gs$, $0.1<y<0.9$, $0.03<x<0.7$, $W>5~\Ge$. This will allow us to directly compare our results for single hadron SSAs with the measurements of Ref.~\cite{Adolph:2012sp}. For the Sivers function we use the functional form of Eq.~(\ref{EQ_SIV_PDF}), with the parameters taken from Ref.~\cite{Anselmino:2005nn}, and slightly adjusted (within their uncertainties) to best reproduce the single hadron SSA measurements. 

 The plots in Fig.~\ref{PLOT_SIV_1H} present the mLEPTO results  for the dependence of the single hadron SIDIS SSAs  on:  (a) the light-cone momentum fraction of the quark $x$, (b) the produced hadrons' light-cone momentum fraction $z$, and (c) the transverse momentum $P_T$, for both positively and negatively charged hadrons. Here we imposed kinematic cuts similar to those used by the COMPASS collaboration~\cite{Adolph:2012sp}: $P_T>0.1~\Ge$ and $z>0.2$. Also depicted in Fig.~\ref{PLOT_SIV_1H} are the COMPASS results of Ref.~\cite{Adolph:2012sp}. We see that mLEPTO reproduces the experimental data well.

\begin{figure}[tb]
\centering 
\subfigure[] {
\includegraphics[width=\ImS]{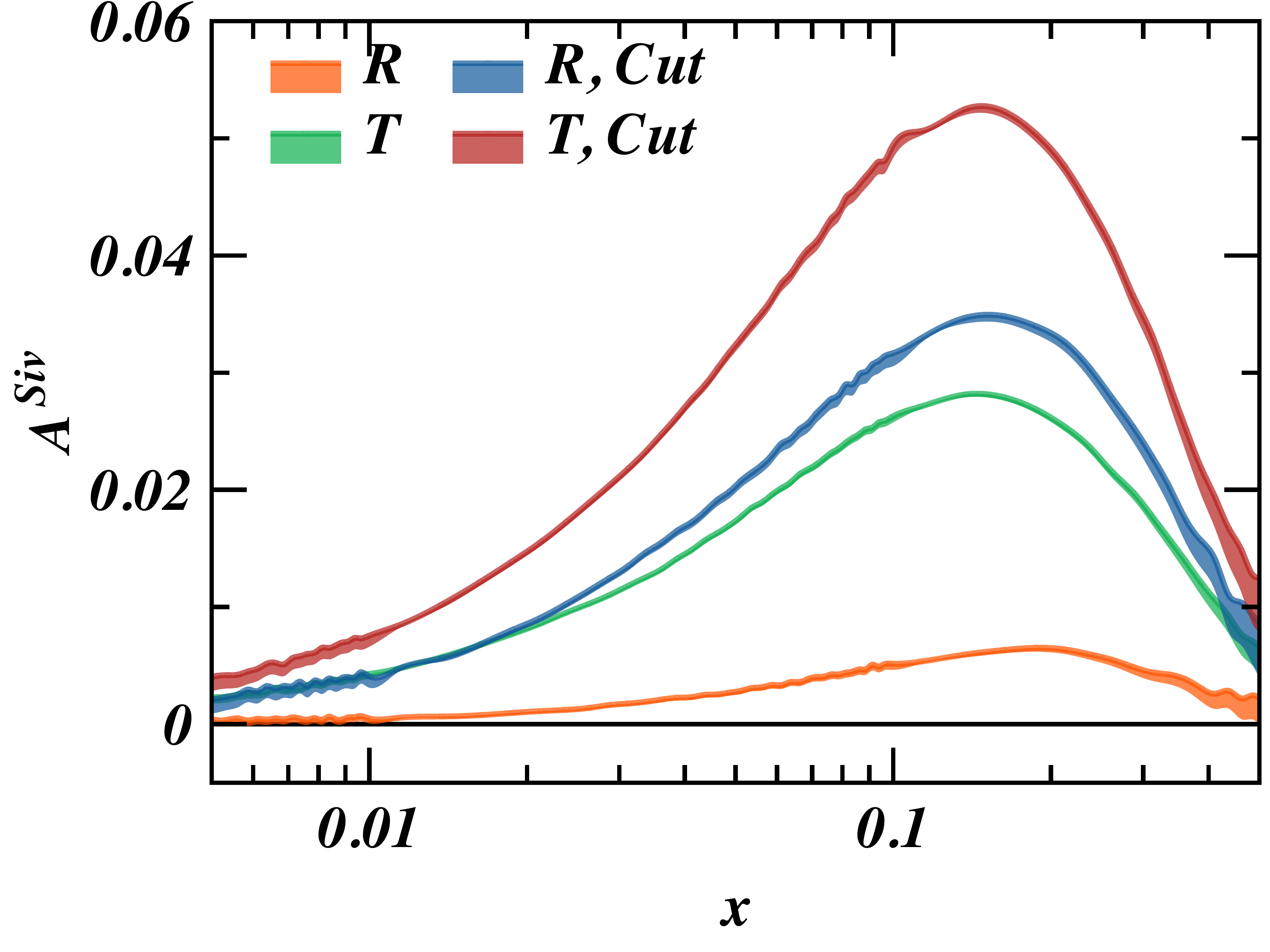}}
\\\vspace{-0.2cm}
\subfigure[] {
\includegraphics[width=\ImS]{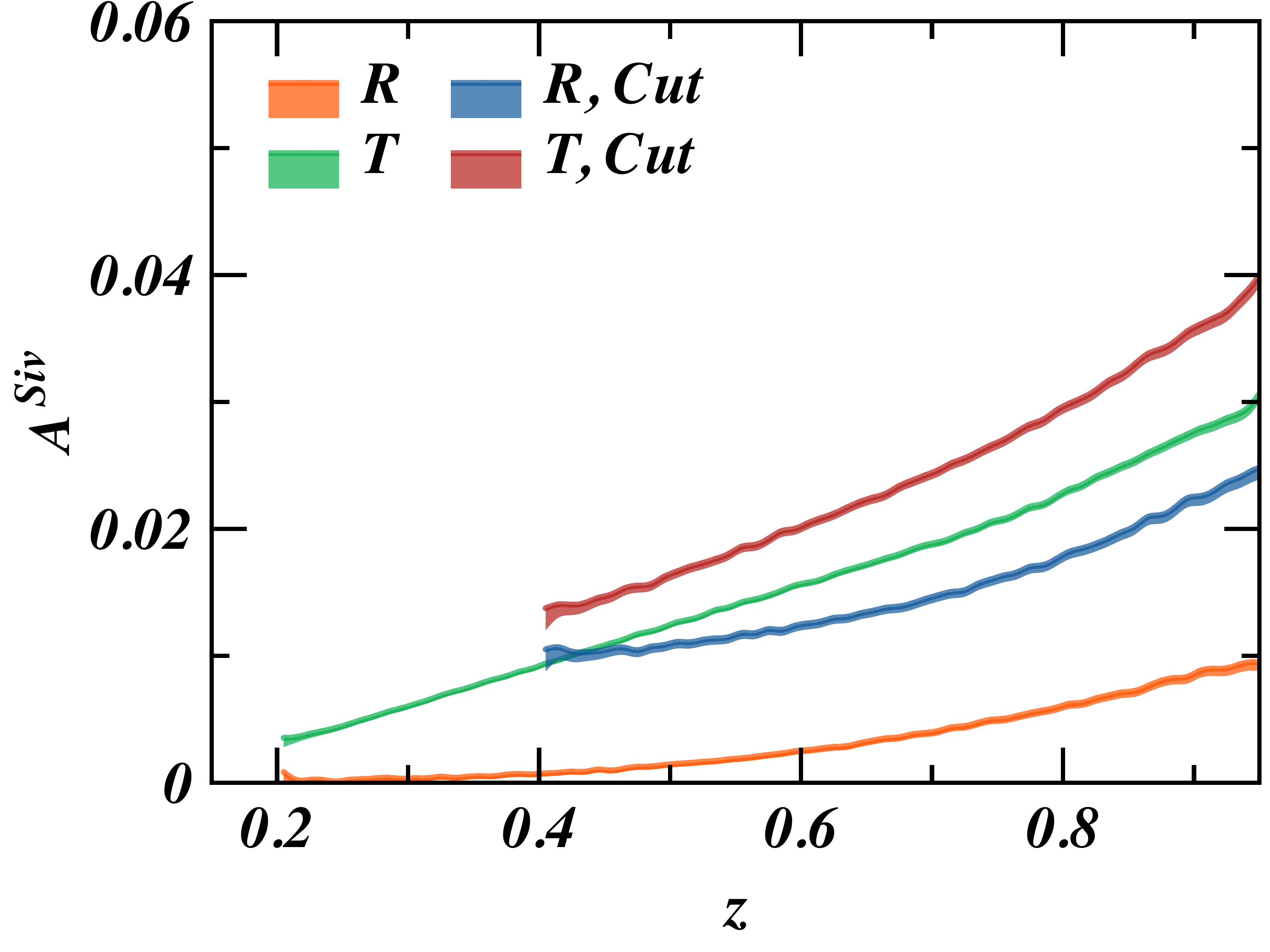}
}
\\\vspace{-0.2cm}
\caption{mLEPTO predictions for the dependence of the Sivers asymmetry on: (a) $x$, and (b) the total light-cone momentum fraction $z$, in oppositely charged hadron pair production off proton target for both $\fR$ and $\fT$ asymmetries integrated over $\vect{P}_T$ and $\vect{R}$, respectively. The lines labeled "Cut" are the results with the additional cut on the positively charged hadron's momentum, as described in the text.}
\label{PLOT_SIV_2H}
\end{figure}

 Figure~\ref{PLOT_SIV_2H} depicts our predictions for the dependence of the Sivers SSAs in two oppositely charged hadron $h^+h^-$ production in SIDIS (the first hadron is taken as $h^+$ and the second as $h^-$) on the quark $x$ in  \ref{PLOT_SIV_2H}(a) and the total light-cone momentum fraction of the pair $z=z_1+z_2$ in \ref{PLOT_SIV_2H}(b). Here, the following kinematic cuts have been imposed $P_{1(2)T}>0.1~\Ge$, $z_{1(2)}>0.1$. In addition, the lines labeled "Cut" depict the results with additional cuts on the momentum of the positively charged hadron: $z_1>0.3$ and $P_{T1}>0.3~\Ge$. These asymmetric cuts significantly enhance the SSAs in all the channels presented.

\vspace{-0.4cm}
{\it Conclusions.}---In this Letter, we demonstrated for the first time that the Sivers SSA in two-hadron semi-inclusive production has a term proportional to $\sin(\fR-\fS)$. We used a simple parametrization of the relevant PDFs and DiFFs to show that the SSAs corresponding to both $\sin(\fR-\fS)$ and $\sin(\fT-\fS)$ modulations are nonvanishing, especially when one imposes asymmetric cuts on the momenta of the hadrons in the pair. In addition, we used the mLEPTO event generator to first describe COMPASS collaboration measurements of the $x$, $z$, and $P_T$ dependence of the SSAs for the Sivers effect in single hadron SIDIS. Then, we used the same kinematics in mLEPTO to calculate for the first time predictions for the two-hadron SSAs depending on $x$ and $z$. While these SSAs are nonzero even with symmetric cuts on the transverse momenta of the hadron pair, asymmetric cuts can significantly enhance the SSAs in all channels. Moreover, the experimental measurement of these SSAs in two-hadron SIDIS would add a valuable input for extracting the Sivers PDFs, alongside the SSAs from single hadron SIDIS and Drell-Yan measurements. Our simulations for the SoLID experiment at JLab12 produce SSAs comparable to those for COMPASS. Detailed derivations of the expressions for the cross sections presented here, as well as predictions for like charged hadron pairs, are presented in our forthcoming article~\cite{Kotzinian:2014siv}.

\vspace{-0.2cm}
We thank M.~Radici and A.~Bacchetta for their comments. A.K. was partially supported by CERN TH division and INFN Torino, H.M. and A.T. were supported by the Australian Research Council through Grants No. FL0992247 (AWT), No. CE110001004 (CoEPP), and by the University of Adelaide. 
\vspace{-0.4cm}
\bibliographystyle{apsrev}
\bibliography{fragment}

\end{document}